\documentclass[prd,amsmath,twocolumn,10pt,superscriptaddress,floatfix,nofootinbib]{revtex4-2}
\usepackage{epsfig,amssymb,amsfonts,amsmath,mathtools,bm,color,xcolor,graphicx,braket,adjustbox,esint,upgreek}
\usepackage{multirow}
\usepackage[utf8]{inputenc}
\usepackage{setspace}

\usepackage{orcidlink}
\usepackage{dcolumn,kantlipsum}
\allowdisplaybreaks

\usepackage{orcidlink}
\usepackage{mathrsfs} 

\usepackage{bm,bbm}
\usepackage{slashed}

\newcommand{\be}{\begin{equation}}
\newcommand{\ee}{\end{equation}}
\newcommand{\bea}{\begin{eqnarray}}
\newcommand{\eea}{\end{eqnarray}}
\newcommand{\beas}{\begin{eqnarray*}}
\newcommand{\eeas}{\end{eqnarray*}}

\renewcommand{\arraystretch}{1.2}
\def\vec#1{\boldsymbol{#1}}
\newcommand{\nn}{\nonumber}

\renewcommand{\L}{\mathscr{L}}

\newcommand{\MeV}{\,\text{MeV}}

\renewcommand{\vec}[1]{\mathbf{#1}}
\newcommand{\diff }{{\text{d}}}

\newcommand{\ustb}{\affiliation{School of Mathematics and Physics, University
of Science and Technology Beijing, Beijing 100083, China}}

\newcommand{\bonn}{\affiliation{Helmholtz-Institut f\"ur Strahlen- und Kernphysik (Theorie) and\\ Bethe Center for Theoretical Physics, Universit\"at Bonn,
53115 Bonn, Germany}}

\newcommand{\julich}{\affiliation{Institute for Advanced Simulation, Forschungszentrum J\"ulich, 52425 J\"ulich, Germany}}

\newcommand{\itp}{\affiliation{Institute of Theoretical Physics, Chinese Academy of Sciences, Beijing 100190, China}}

\newcommand{\ucas}{\affiliation{School of Physical Sciences, University of Chinese Academy of Sciences, Beijing 100049, China}}

\newcommand{\scnt}{\affiliation{Southern Center for Nuclear-Science Theory (SCNT), Institute of Modern Physics,\\ Chinese Academy of Sciences, Huizhou 516000, China}}

\graphicspath{{FiguresV2/}}

\begin{document}

\title{Decoding the structure near the \boldmath{$\pi^+\pi^-$ mass threshold in $\psi(3686) \rightarrow
J/\psi \pi^+\pi^-$} decays}

\author{Yun-Hua Chen\orcidlink{0000-0001-8366-2170}}\email{ yhchen@ustb.edu.cn}
\ustb

\author{Xiang-Kun Dong\orcidlink{0000-0001-6392-7143}}\email{ xiangkun@hiskp.uni-bonn.de}
\bonn

\author{Feng-Kun Guo\orcidlink{0000-0002-2919-2064}}\email{ fkguo@itp.ac.cn}
\itp \ucas \scnt

\author{Christoph Hanhart\orcidlink{0000-0002-3509-2473}}\email{c.hanhart@fz-juelich.de }            
\julich

\author{Bastian Kubis\orcidlink{0000-0002-1541-6581}}\email{kubis@hiskp.uni-bonn.de }
\bonn

\begin{abstract}

In light of recent high-precision data taken by the BESIII Collaboration, we reconsider the dipion transition $\psi(3686) \rightarrow J/\psi \pi^+\pi^-$. The strong pion-pion final-state interactions are taken into account model independently by using dispersion theory. 
We find that we can reproduce the substructure near the $\pi^+\pi^-$ threshold observed experimentally without introducing an extra resonance state. While a helicity-flip amplitude plays an important role for the formation of the dip in the invariant-mass distribution, the virtual exchange of the charmoniumlike exotic $Z_c(3900)$ state improves the fit quality only slightly.

\end{abstract}


\maketitle

\newpage

\section{Introduction}

The dipion transitions between heavy quarkonia are important for understanding both their dynamics and low-energy QCD. Because the heavy quarkonia
are expected to be nonrelativistic and compact, the method of the
QCD multipole expansion~\cite{Voloshin:1980zf,Novikov:1980fa,Kuang:1981se} is
often used to study these transitions, in which the pions are produced due to the hadronization of emitted soft gluons.  
The QCD multipole expansion maps onto the chiral effective Lagrangian for dipion transitions between heavy quarkonia~\cite{Brown:1975dz, Mannel:1995jt}, which has been used successfully together with subsequent pion-pion final-state interactions (FSI) to study the decays $\Upsilon(2S,3S,4S,5S) \to \Upsilon(1S,2S,3S)\pi\pi$~\cite{Chen:2015jgl,Chen:2016mjn,Chen:2019gty,Baru:2020ywb}
and $\psi(3686) \to J/\psi \pi\pi$~\cite{Chen:2019hmz,Dong:2021lkh}.

Very recently, based on $2712.4(14.4)\times 10^{6}$ $\psi(3686) \equiv \psi'$ events, BESIII presented a high-precision study of the $\pi^+\pi^-$ mass spectrum in $\psi'\rightarrow\pi^{+}\pi^{-}J/\psi$ decays. 
A clear resonancelike structure near the $\pi^+\pi^-$ mass threshold was observed for the first time~\cite{BESIII:2025ozb}. In the BESIII analysis, the structure was best described by a Breit-Wigner parametrization with a mass of $(282.6 \pm 0.4_{\mathrm{sta}} \pm 2.5_{\mathrm{sys}})\,$MeV and a width of $(17.3 \pm 0.8_{\mathrm{sta}} \pm 0.4_{\mathrm{sys}})\,$MeV~\cite{BESIII:2025ozb}. 
However, a resonance with such a low mass, which is much lower than the mass of the $f_0(500)$ (for a review, see Ref.~\cite{Pelaez:2015qba}), would be at odds with the chiral structure of QCD that suppresses two-pion interactions at low energies. Thus, the new dataset necessitates a reanalysis of the reaction $\psi^\prime \to J/\psi \pi\pi$. 
Note that the dipion mass spectra in $\Upsilon(3S) \to \Upsilon(1S) \pi \pi $, which show a similar two-hump behavior, can be  explained naturally by the virtual exchange
of the isovector bottomonium exotics $Z_b(10610)/Z_b(10650)$~\cite{Chen:2015jgl} (see Refs.~\cite{Anisovich:1995zu,Guo:2004dt} for earlier studies), and similar double-bump structures were also observed in $\Upsilon(4S) \to \Upsilon(1S,2S) \pi\pi$ decays as discussed in Ref.~\cite{Guo:2006ai}.
The analogous isovector charmoniumlike  structure $Z_c(3900)^\pm$
was discovered in the $J/\psi\,\pi$ invariant-mass spectra by the BESIII and Belle Collaborations in 2013 in the process $e^+e^-\to  J/\psi\pi^+\pi^-$~\cite{BESIII:2013ris,Belle:2013yex}.
Therefore, we study the decay $\psi^\prime \to J/\psi \pi\pi$, considering effects of the virtual
$Z_c(3900)$-exchange mechanism.

It is necessary to account for the 
two-pion FSI properly. In this article, we  use dispersion theory to treat the $\pi\pi$ FSI~\cite{Garcia-Martin:2010kyn,Kubis:2015sga,Kang:2013jaa,Dai:2014lza,Chen:2015jgl}. Instead of the chiral unitary approach~\cite{Oller:1997ti} that was used in the literature to describe $\pi\pi$ FSI for the process of interest~\cite{Guo:2004dt,Guo:2006ya}, in dispersion theory it is treated in a model-independent way consistent with $\pi\pi$ scattering data. We will provide a simultaneous description of the experimental data for the dipion invariant-mass distribution and the helicity angular
distribution of $\psi' \to J/\psi \pi\pi$, which help reveal the nature of the structure near the $\pi^+\pi^-$ threshold.

\section{Theoretical framework}
\label{theor}

For the contact $\psi'\to J/\psi\pi\pi$ interaction, the effective
Lagrangian to leading order in the chiral as well as the heavy-quark
nonrelativistic expansion reads~\cite{Mannel:1995jt,Chen:2015jgl,Chen:2016mjn,Yan:2026oil} 
\begin{align}
\L_{\psi^\prime\psi\Phi\Phi} =& g_1\langle \psi^{\prime\alpha} \psi^\dag_\alpha \rangle \langle u_\mu
u^\mu\rangle +h_1\langle \psi^{\prime\alpha} \psi^\dag_\alpha \rangle \langle u_\mu u_\nu\rangle
v^\mu v^\nu \nn\\ &+j_1\langle \psi^{\prime\mu} \psi^\dag_\nu \rangle \langle u_\mu u^\nu\rangle
+c_m\langle \psi^{\prime\mu} \psi^\dag_\mu \rangle \langle \chi_+\rangle
+\mathrm{H.c.}\,, \label{LagrangianYpsipipi}
\end{align}
where $v^\mu=(1,\vec{0})$ is the velocity of the heavy quark,
and $\psi^{\alpha}$  and $\psi^{\prime\alpha}$ denote the $J/\psi$ and 
$\psi'$ fields, respectively.
$u_\mu$ collects the Goldstone bosons of the spontaneous breaking of
chiral symmetry~\cite{Gasser:1983yg,Gasser:1984gg},
\bea
u_\mu &=& i \left( u^\dagger \partial_\mu u\, -\, u \partial_\mu u^\dagger\right) \,, \qquad
u^2 = e^{i {\Phi}/{ F_\pi}}\,, \qquad \nn\\
\Phi &=&
\begin{pmatrix}
   \pi ^0  & \sqrt2{\pi^+ }  \\
   \sqrt2{\pi^- } & -\pi ^0  \\
\end{pmatrix} ,
\eea
where $F_\pi=92.2\MeV$ is the pion decay constant, $\chi_+ = u^\dagger   \chi u^\dagger + u \chi^\dagger u$,
with $\chi=2 B \, {\rm diag}(m_u,m_d)$, and $B$ is a constant related to the quark condensate in the chiral limit.
The leading-order 
$Z_c \psi^\prime\pi$ and $Z_c J/\psi\pi$ interaction Lagrangians
are proportional to the pion energy~\cite{Cleven:2011gp},
\begin{align}
\L_{Z_c\psi^\prime\pi} &= C_{Z_c \psi^\prime\pi} \psi^{\prime i} \langle {Z^i_{c}}^\dagger u_\mu 
v^\mu \rangle +\mathrm{H.c.}\,, \notag\\
\L_{Z_c\psi\pi} &= C_{Z_c \psi\pi} \psi^i \langle {Z^i_{c}}^\dagger u_\mu 
v^\mu  \rangle +\mathrm{H.c.}
 \,. \label{LagrangianZc}
\end{align}
We define $C_{\psi^\prime\psi}^{Z_c}\equiv C_{Z_c \psi^\prime\pi} C_{Z_c \psi\pi}$ as the product
of the coupling constants for the exchange of the $Z_{c}$.

The doubly differential decay width for the process $\psi'(p_a) \rightarrow J/\psi(p_b)\pi(p_c)\pi(p_d)$ can be written as
\begin{align}
\frac{\diff^2 \Gamma}{\diff \sqrt{s} \,\diff \cos\theta}    = \frac{\sqrt{s}\kappa(s)}{32(2\pi)^3 M_{\psi^\prime}^3} \frac{1}{3}\sum_{\lambda, \lambda^\prime} \left|\mathscr{M}^{\lambda,\lambda^\prime}(s,t,u)\right|^2\,,
\label{eq.DifDecayWidthJpsipipi}
\end{align}
where the Mandelstam variables are
$s=(p_c+p_d)^2$, $t=(p_a-p_c)^2$, and $u=(p_a-p_d)^2$, and $\theta$ is defined as the angle between $\pi^+$ and $\psi'$ in the $\pi\pi$ c.m.\ frame,
$\kappa(s)=\sqrt{1-4m_\pi^2/s}\lambda^{1/2}(M_{\psi^\prime}^2,s,M_\psi^2)$, and
$\lambda(a,b,c)=a^2+b^2+c^2-2(ab+ac+bc)$ is the K\"all\'en triangle function. 
$\mathscr{M}^{\lambda,\lambda^\prime}(s,t,u)\equiv \mathscr{M}_{\mu\nu}(s,t,u)\epsilon_{\lambda^\prime}^\mu \epsilon_{\lambda}^{\ast\nu}$ is the helicity amplitude, and $\lambda^\prime$ and $\lambda$ denote the helicities of $\psi^\prime$ and $J/\psi$, respectively.
To proceed, we perform the $s$-channel partial-wave decomposition of the amplitude
\begin{align}\label{eq.PWHelicityAmp}
\mathscr{M}^{\lambda,\lambda^\prime}(s,t,u)  &= \sum_{l=0}^{\infty}   d_{\lambda-\lambda^\prime,0}^l(\theta) S_l^{{1}/{2}}(s)\nn\\
&\quad \times\left[H^{\lambda, \lambda^\prime,l}(s)+\hat{H}^{\lambda, \lambda^\prime,l}(s)\right],
\end{align}
where $l$ denotes the relative orbital angular momentum of the pions, and $d_{\lambda-\lambda^\prime,0}^l(\theta)$ is the Wigner-$d$ function.
In order to take the near-threshold Coulomb enhancement of the $\pi^+\pi^-$ pair into account, we employ the Sommerfeld factors~\cite{Sommerfeld:1931qaf,Abramowitz,Cassel:2009wt} 
\begin{align}
     S_0(s)&= \frac{2\pi x}{1-\text{exp}(-2\pi x)}, \nn\\
     S_{l>0}(s)&= S_0(s)\times \prod_{b=1}^{l}\Big(1+\frac{x^2}{b^2}\Big),
\end{align} 
with $ x =  {\alpha \, m_\pi}/{\sqrt{s-4m_\pi^2}}$,
where $\alpha = e^2/(4\pi)\approx 1/137$ is the fine-structure constant.
$H^{\lambda, \lambda^\prime,l}(s)$ is
the right-hand-cut part and represents the $s$-channel $\pi\pi$ rescattering. The ``hat function'' $\hat{H}^{\lambda, \lambda^\prime,l}(s)$ encodes the left-hand-cut contribution
derived from the partial-wave projection of the $Z_c$-exchange amplitude obtained from Eq.~\eqref{LagrangianZc}. Charge-parity conservation requires that $l$ must be even, and we only take the $S$- and $D$-wave components into account in this study.

In the regime of elastic $\pi\pi$ rescattering, the partial-wave unitarity relations read
\begin{align}\label{eq:disc}
\text{Im}H^{\lambda, \lambda^\prime,l}(s) =
\big[H^{\lambda, \lambda^\prime,l}(s)+\hat{H}^{\lambda, \lambda^\prime,l}(s)\big] \sin\delta_l^0(s)e^{-i \delta_l^0(s)}\,,
\end{align}
where $\delta_l^0(s)$ is the $\pi\pi$ isoscalar phase shift of angular momentum $l$.

\begin{figure}
\includegraphics[width=\linewidth]{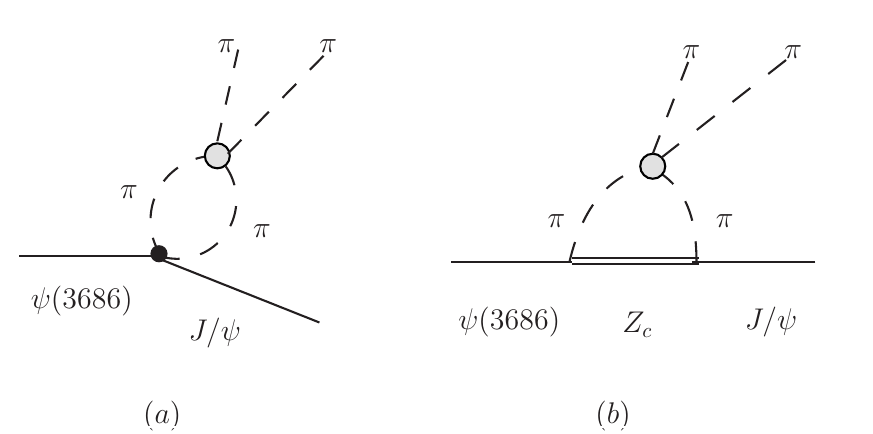}
\caption{Diagrams considered for $\psi' \rightarrow
J/\psi \pi^+\pi^-$. Diagram (a) denotes the contributions of the chiral contact terms, while diagram (b) corresponds to the contributions of the $Z_c$ exchange. The gray blob denotes
pion-pion rescattering.} \label{fig:FeynmanDiagram}
\end{figure}

The dispersive solution to Eq.~\eqref{eq:disc} is~\cite{Anisovich:1996tx}
\begin{align}\label{OmnesSolution1channel}
H^{\lambda, \lambda^\prime,l}(s)=&\,\Omega_l^0(s)\bigg\{M^{\chi,\lambda,\lambda^\prime,l}(s)   \nn \\& +\frac{s^n}{\pi}\int_{4m_\pi^2}^\infty
\frac{\diff x}{x^n}\frac{\hat{H}^{\lambda, \lambda^\prime,l}(x)\sin\delta_l^0(x)}{|\Omega_l^0(x)|(x-s)}\bigg\} \,,
\end{align}
where the subtraction terms $M^{\chi,\lambda,\lambda^\prime,l}(s)$ are matched to the partial-wave projection of the low-energy chiral amplitudes obtained from Eq.~\eqref{LagrangianYpsipipi}.
The single-channel Omn\`es function reads~\cite{Omnes:1958hv}
\begin{equation}\label{Omnesrepresentation}
\Omega_l^0(s)=\exp
\bigg\{\frac{s}{\pi}\int^\infty_{4m_\pi^2}\frac{\diff x}{x}
\frac{\delta_l^0(x)}{x-s}\bigg\}\,.
\end{equation}
For the $S$ wave, we use the phase of the nonstrange pion
scalar form factor as determined in Ref.~\cite{Hoferichter:2012wf}. For the $D$ wave,
we employ the parametrization for $\delta_2^0$ given in 
Ref.~\cite{Garcia-Martin:2011iqs}. A diagrammatic representation of all contributions is
given in Fig.~\ref{fig:FeynmanDiagram}.

\section{Phenomenological discussion} \label{pheno}

In this work, we perform fits taking into account the experimental datasets of the $\pi\pi$
invariant-mass distributions and the helicity angular distribution for $\psi' \to J/\psi \pi^+\pi^-$ measured by the BESIII Collaboration~\cite{BESIII:2025ozb}.

\begin{figure}
\centering
\includegraphics[width=\linewidth]{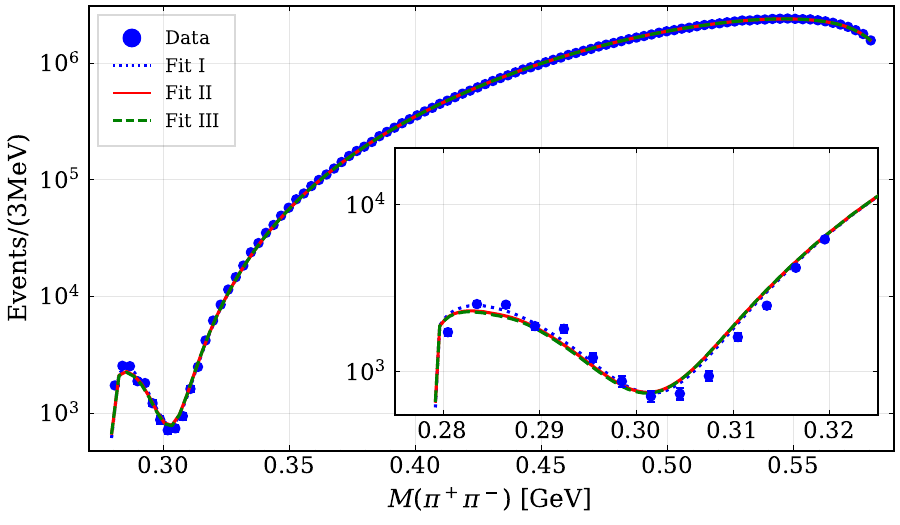}
\includegraphics[width=\linewidth]{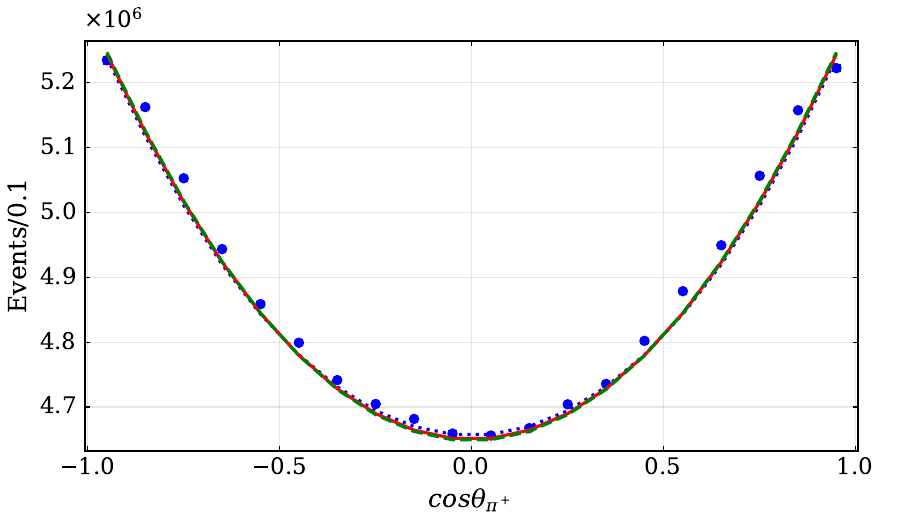}
\caption{Fit results of the $\pi\pi$ invariant-mass spectra (top) and the $\cos\theta$ distributions (bottom) in $\psi' \to J/\psi \pi\pi $ of Fits I (blue dotted), II (red solid), and III (green dashed).  The enlarged inset (top) shows the fit results in the region of $(0.28, 0.32)$~GeV. In the $\cos\theta$ distributions (bottom), the zero is suppressed to make differences in the fits visible. The data are taken from Ref.~\cite{BESIII:2025ozb}.  }\label{fig.MassAngularDistributions}
\end{figure}

\begin{figure}
\centering
\includegraphics[width=\linewidth]{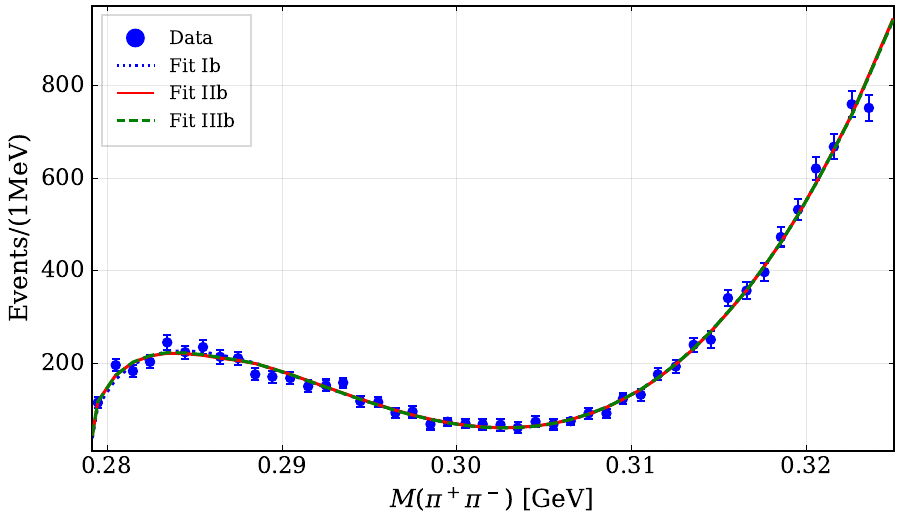}
\caption{Fit results of the folded $\pi\pi$ spectra data without efficiency correction in the range of $M(\pi^+\pi^-) \in (0.28, 0.32)$~GeV in $\psi' \to J/\psi
\pi\pi $ of Fits Ib (blue dotted), IIb (red solid), and IIIb (green dashed). } \label{fig.MassDistributionsLowEnergy}
\end{figure}

To illustrate the effect of the $Z_c$-exchange term, we perform three fits. 
In Fit I, we only consider the contribution of the chiral contact terms, while in Fits~II and~III, the $Z_c$ exchange is taken into account in addition. In Fit II, the mass and width of $Z_c$ are fixed at the central values determined in our previous work~\cite{Chen:2023def}: $M_{Z_c} = 3.880$~GeV and $\Gamma_{Z_c}= 0.036$~GeV. In contrast, in Fit III we treat the mass and width of $Z_c$ as free parameters within the range obtained in Ref.~\cite{Chen:2023def}: $M_{Z_c} \in 3.880(24)$~GeV and $\Gamma_{Z_c} \in 0.036(17)$~GeV.
The fitted results of Fits~I,~II, and~III are shown as the blue dotted, red solid, and green dashed lines in Fig.~\ref{fig.MassAngularDistributions}, respectively.
The fitted parameters as well as the $\chi^2$ per degree of freedom ($\chi^2/\text{dof}$) are shown in Table~\ref{tablepar1}. As can be seen from Fig.~\ref{fig.MassAngularDistributions} and Table~\ref{tablepar1}, all the Fits~I,~II, and~III agree with the experimental data in most regions by eye, while their ${\chi^2}/{\rm dof}$ are much larger than 1. 
Note that the BESIII paper~\cite{BESIII:2025ozb} alternatively presents the folded $\pi\pi$ spectra data in the low-energy range of $M(\pi^+\pi^-) \in (0.28, 0.32)$ GeV, which were fitted using a Breit-Wigner function convolved with a Gaussian energy resolution function and multiplied by the detection efficiency. 
Therefore, we perform three further Fits Ib, IIb, and IIIb for the folded $\pi\pi$ spectra dataset, neglecting the detection efficiency corrections, using the same strategy as in Fits I, II, and III, respectively. The result is shown in Fig.~\ref{fig.MassDistributionsLowEnergy}, and the ${\chi^2}/{\rm dof}$ values for Fits Ib, IIb, and IIIb are 1.04, 1.03, and 1.08, respectively.

In our Fits~I and~Ib, the enhancement near the $\pi\pi$ threshold can be described by the chiral contact terms and FSI.
In particular, no additional 
2$\pi$ resonance is required. Note that Ref.~\cite{BESIII:2025ozb} claims that chiral perturbation theory supplemented with a leading-order unitarization as used in Ref.~\cite{Guo:2004dt} fails to reproduce this threshold enhancement. We find that in Ref.~\cite{BESIII:2025ozb}, the $E_{\pi^+}E_{\pi^-}$ and $(p_{\pi^+}^{\mu}p_{\pi^-}^{\nu}+p_{\pi^+}^{\nu}p_{\pi^-}^{\mu})$ parts are considered as pure $\pi\pi$ $S$ and $D$ waves, respectively, where $E_{\pi^{(\pm)}}$ and $p_{\pi^{(\pm)}}$ are the energies and the four-momenta of the pions, while both of these actually contribute to both $S$ and $D$ waves~\cite{Yan:1998wv,Guo:2004dt}. Also note that Ref.~\cite{BESIII:2025ozb} considers the FSI within the chiral unitary approach~\cite{Oller:1997ti}. 
While it already yields a dynamical generation of the scalar mesons, the very precise information available on 
pion-pion phase shifts is not strictly implemented there.

Comparing Fit~I with Fits~II and~III, one finds that including the $Z_c$ exchange improves the fit quality, especially for the helicity angular distribution. Furthermore, comparing Fits~II and~III, one observes that although a smaller $Z_c$ mass leads to a smaller $\chi^2$, the change of fit quality is tiny. Therefore we conclude that a virtual intermediate $Z_c$ state indeed can play a relevant role in $\psi' \to J/\psi \pi^+\pi^-$.

\begin{table}
\caption{\label{tablepar1}   Fit parameters from the simultaneous fit of the efficiency-corrected unfolded $\pi\pi$
invariant-mass distributions and the helicity angular distribution for the $\psi' \to J/\psi
\pi^+\pi^-$ process. Fit~I: contributions of the chiral contact terms only.
Fits~II and~III: contributions of the chiral contact terms and the $Z_c$ exchange. In Fit II, the mass and width of the $Z_c$ are fixed at $M_{Z_c} = 3.880$~GeV and $\Gamma_{Z_c}= 0.036 $~GeV, while in Fit III, they are treated as free parameters within the ranges $M_{Z_c} \in 3.880(24)$~GeV and $\Gamma_{Z_c} \in 0.036(17)$~GeV~\cite{Chen:2023def}. 
The $\chi^2$ values are explicitly presented as the sum of contributions from the $\pi\pi$ invariant-mass distribution (the first number) and the helicity angular distribution (the second number).
}
\renewcommand{\arraystretch}{1.3}
\begin{center} 
   \begin{ruledtabular}
\begin{tabular}{l|ccc} 
        & Fit~I
         & Fit~II
         & Fit~III \\
\hline
$g_1~(\text{GeV}^{-1})$   &    $ 2.972(1)$  &    $ 3.368(1)$&    $ 3.456(1)$\\
$h_1~(\text{GeV}^{-1})$   &    $ -1.680(6)$  &    $ -2.435(2)$ &    $ -2.571(3)$ \\
$j_1~(\text{GeV}^{-1})$   &    $ -0.281(3)$  &   $ -0.260(3)$&   $ -0.274(3)$\\
$c_m~(\text{GeV}^{-1})$   &    $ 2.112(23)$  &   $ 0.424(7)$&   $ -0.012(7)$\\
$C_{\Psi^\prime\Psi}^{Z_c}$  &    $\ast$  &     $ -1.217(3)$ &     $ -1.443(5)$ \\
$M_{Z_c}~(\text{GeV})$  &   $\ast$  &$ 3.880 ~\text{(fixed)}$  &$ 3.873(2)$ \\
$\Gamma_{Z_c}~(\text{GeV})$  &    $\ast$  &    $ 0.036 ~\text{(fixed)}$ &    $ 0.019(1)$ \\
\hline
 ${\chi^2}/{\rm dof}$ &  $\frac{1105+479}{(121-4)}$  &  $\frac{1095+327}{(121-5)}$  &  $\frac{1086+289}{(121-7)}$   \\
 & $=13.5$ & $=12.3$ & $=12.1$  
\end{tabular}
\end{ruledtabular}
\end{center}
\renewcommand{\arraystretch}{1.0}
\end{table}

\begin{figure}
\centering
\includegraphics[width=\linewidth]{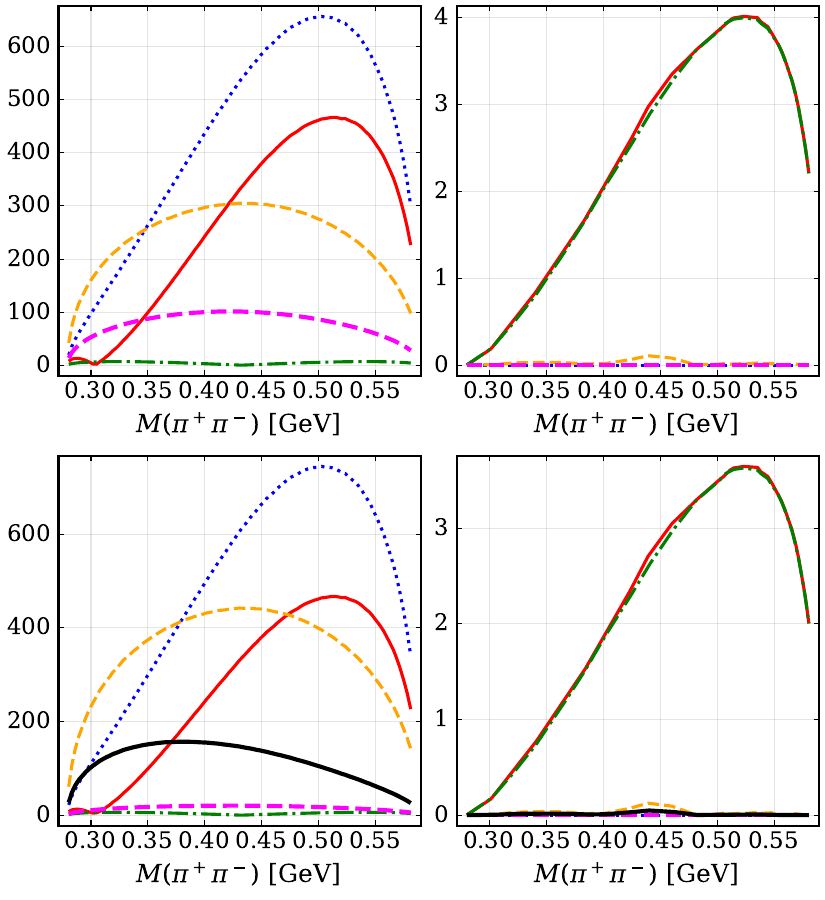}
\caption{Moduli of the $S$- (left) and $D$-wave (right)
amplitudes for $\psi' \to J/\psi \pi^+\pi^-$ in
Fits~I (top) and II (bottom). The red solid lines represent our best fit results,
while the blue dotted, orange dashed,
dark green dot-dashed, thick magenta dashed, and thick black solid lines correspond to the contributions
from the chiral contact terms proportional to $g_1$, $h_1$, $j_1$, and $c_m$, as well as the $Z_c$-exchange term, respectively.
}
\label{fig.Moduli}
\end{figure}

In Fig.~\ref{fig.Moduli}, we plot the moduli of the $S$- and
$D$-wave amplitudes from the chiral contact terms proportional to $g_1$, $h_1$, $j_1$, $c_m$, and the $Z_c$-exchange term for Fits~I and II.
We observe that the $D$-wave amplitudes are about two orders of magnitude smaller than the corresponding $S$-wave amplitudes. Therefore, a flat
angular distribution is expected in our scheme, which agrees with the BESIII measurement. For the $S$-wave amplitudes, the contribution from the helicity-flip term is much smaller than the others in most regions as expected. While the contribution from the $j_1$ term has a zero in the energy region of interest, its effect is non-negligible for the low-energy dip in the $\pi\pi$ invariant-mass distribution.
Also note that for the $S$-wave amplitudes, the shapes of the curves corresponding to the $c_m$ term as shown in Fit~I (upper left panel of Fig.~\ref{fig.Moduli}) 
and the $Z_c$-exchange term as shown in Fit~II (lower left panel of Fig.~\ref{fig.Moduli}) are too similar to be 
disentangled by a fit to the $\pi\pi$ invariant mass distribution. Nevertheless, we find that either the $c_m$ term or the $Z_c$-exchange term is indispensable for describing the experimental data well.
For the $D$-wave amplitudes, the helicity-flip contribution dominates by far. This means that the curved behavior of the observed angular distribution is mainly due to the $j_1$ term. We find that the $D$-wave contribution from the $Z_c$ exchange is much smaller than its $S$-wave amplitude, which is due to the fact that the $Z_c$ is virtual in this process.

\section{Conclusions}
\label{conclu}

We used dispersion theory to study the FSI in the decay
$\psi' \to J/\psi \pi\pi$. Through fitting the data of the $\pi\pi$ mass spectra and
the angular $\cos\theta$ distributions, we find that the structure near the $\pi^+\pi^-$ mass threshold observed by~\cite{BESIII:2025ozb} can be described well without introducing a new resonance. We find that the helicity-flip amplitude plays an important role in the low-energy $\pi\pi$ mass spectrum and the angular distributions. Also we find that the virtual $Z_c(3900)$-exchange mechanism improves the agreement between the theoretical prediction and the experimental data.

Note added: during completion of this manuscript we
became aware of Ref.~\cite{Wang:2025kcy} addressing the
 same problem and explaining the structure near threshold as the result from the
interference and superposition of two decay mechanisms. We point out that the second decay mechanism in Fig.~1 of Ref.~\cite{Wang:2025kcy} in practice
refers to a contribution for Goldstone-boson singlets, i.e., it
exclusively gives a contribution to the $\eta\eta$ channel,
for which they need to include non-octet parts of the $\eta$. In every
other coupled-channel analysis of the $I=0$ $S$ wave, the $\eta\eta$ channel is
tiny compared to the $\pi\pi$--$K\bar{K}$ channel coupling~\cite{Kaiser:1998fi,Garcia-Martin:2011iqs,Briceno:2017qmb}; indeed, from the fit results in Table I of Ref.~\cite{Wang:2025kcy}, in order for this mechanism
to have any effect, the corresponding coupling $V_2$ is found to be larger by five orders of magnitude than that for the octet meson pair. Also, the complex phase $\phi$, although very small, should not exist required by the quantum field theory.  

\section*{Acknowledgments}

We are grateful to Shuang-shi Fang and Ben-Hou Xiang for helpful discussions and for providing the BESIII data, and we thank Ulf-G.\ Mei\ss{}ner for helpful discussions. This work is supported in part by the Fundamental Research Funds
for the Central Universities under Grant No.~FRF-BR-19-001A; by the National Natural Science Foundation of China (NSFC) under Grants No.~12125507, No.~12361141819, and No.~12447101; by the National Key R\&D Program of China under Grant No.~2023YFA1606703; and by the Chinese Academy of Sciences (CAS) under Grant No.~YSBR-101. C.H. also acknowledges the support from the CAS President’s International Fellowship Initiative (PIFI) under Grant No.~2025PD0087.

\bibliography{refs}

@article{Voloshin:1980zf,
    author = "Voloshin, Mikhail B. and Zakharov, Valentin I.",
    title = "{Measuring QCD Anomalies in Hadronic Transitions Between Onium States}",
    reportNumber = "DESY-80-28",
    doi = "10.1103/PhysRevLett.45.688",
    journal = "Phys. Rev. Lett.",
    volume = "45",
    pages = "688",
    year = "1980"
}

@article{Abramowitz,
    author = "M.~Abramowitz and Irene A.~Stegun",
    title = "{Handbook of Mathematical Functions with Formulas, Graphs, and Mathematical Tables}",
    journal = "Dover Publications",
    year = "1964"
}

@article{Cassel:2009wt,
    author = "Cassel, S.",
    title = "{Sommerfeld factor for arbitrary partial wave processes}",
    eprint = "0903.5307",
    archivePrefix = "arXiv",
    primaryClass = "hep-ph",
    reportNumber = "OUTP-0910P",
    doi = "10.1088/0954-3899/37/10/105009",
    journal = "J. Phys. G",
    volume = "37",
    pages = "105009",
    year = "2010"
}

@article{Novikov:1980fa,
    author = "Novikov, V. A. and Shifman, Mikhail A.",
    title = "{Comment on the $\psi' \to J/\psi \pi \pi$ Decay}",
    reportNumber = "ITEP-93-1980",
    doi = "10.1007/BF01429829",
    journal = "Z. Phys. C",
    volume = "8",
    pages = "43",
    year = "1981"
}

@article{Kuang:1981se,
    author = "Kuang, Yu-Ping and Yan, Tung-Mow",
    title = "{Predictions for Hadronic Transitions in the $b\bar b$ System}",
    reportNumber = "CLNS-81/495",
    doi = "10.1103/PhysRevD.24.2874",
    journal = "Phys. Rev. D",
    volume = "24",
    pages = "2874",
    year = "1981"
}

@article{Chen:2019hmz,
    author = "Chen, Yun-Hua",
    title = "{Chromopolarizability of Charmonium and {\ensuremath{\pi}}{\ensuremath{\pi}} Final State Interaction Revisited}",
    eprint = "1901.04126",
    archivePrefix = "arXiv",
    primaryClass = "hep-ph",
    doi = "10.1155/2019/7650678",
    journal = "Adv. High Energy Phys.",
    volume = "2019",
    pages = "7650678",
    year = "2019"
}

@article{Dong:2021lkh,
    author = "Dong, Xiang-Kun and Baru, Vadim and Guo, Feng-Kun and Hanhart, Christoph and Nefediev, Alexey and Zou, Bing-Song",
    title = "{Is the existence of a $J/\psi J/\psi$ bound state plausible?}",
    eprint = "2107.03946",
    archivePrefix = "arXiv",
    primaryClass = "hep-ph",
    doi = "10.1016/j.scib.2021.09.009",
    journal = "Sci. Bull.",
    volume = "66",
    number = "24",
    pages = "2462--2470",
    year = "2021"
}

@preprint{BESIII:2025ozb,
    author = "Ablikim, Medina and others",
    collaboration = "BESIII",
    title = "{Observation of a resonance-like structure near the $\pi^+\pi^-$ mass threshold in $\psi(3686) \rightarrow \pi^{+}\pi^{-}J/\psi$}",
    eprint = "2509.23761",
    archivePrefix = "arXiv",
    primaryClass = "hep-ex",
    month = "9",
    year = "2025"
}

@article{BESIII:2013ris,
    author = "Ablikim, M. and others",
    collaboration = "BESIII",
    title = "{Observation of a Charged Charmoniumlike Structure in $e^+e^- \to \pi^+\pi^- J/\psi$ at $\sqrt{s}$ =4.26 GeV}",
    eprint = "1303.5949",
    archivePrefix = "arXiv",
    primaryClass = "hep-ex",
    doi = "10.1103/PhysRevLett.110.252001",
    journal = "Phys. Rev. Lett.",
    volume = "110",
    pages = "252001",
    year = "2013"
}

@article{Garcia-Martin:2010kyn,
    author = "Garc\'ia-Mart\'in, R. and Moussallam, B.",
    title = "{MO analysis of the high statistics Belle results on $\gamma\gamma\to \pi^+\pi^-,\pi^0\pi^0$ with chiral constraints}",
    eprint = "1006.5373",
    archivePrefix = "arXiv",
    primaryClass = "hep-ph",
    doi = "10.1140/epjc/s10052-010-1471-7",
    journal = "Eur. Phys. J. C",
    volume = "70",
    pages = "155--175",
    year = "2010"
}

@article{Oller:1997ti,
    author = "Oller, J. A. and Oset, E.",
    title = "{Chiral symmetry amplitudes in the S wave isoscalar and isovector channels and the $\sigma$, $f_0(980)$, $a_0(980)$ scalar mesons}",
    eprint = "hep-ph/9702314",
    archivePrefix = "arXiv",
    doi = "10.1016/S0375-9474(97)00160-7",
    journal = "Nucl. Phys. A",
    volume = "620",
    pages = "438--456",
    year = "1997",
    note = "[Erratum: Nucl. Phys. A \textbf{652}, 407 (1999)]"
}

@article{Mannel:1995jt,
    author = "Mannel, Thomas and Urech, Res",
    title = "{Hadronic decays of excited heavy quarkonia}",
    eprint = "hep-ph/9510406",
    archivePrefix = "arXiv",
    reportNumber = "TTP-95-36",
    doi = "10.1007/s002880050344",
    journal = "Z. Phys. C",
    volume = "73",
    pages = "541--546",
    year = "1997"
}

@article{Gasser:1983yg,
    author = "Gasser, J. and Leutwyler, H.",
    title = "{Chiral Perturbation Theory to One Loop}",
    reportNumber = "CERN-TH-3689",
    doi = "10.1016/0003-4916(84)90242-2",
    journal = "Annals Phys.",
    volume = "158",
    pages = "142",
    year = "1984"
}

@preprint{Wang:2025kcy,
    author = "Wang, Zhong-Yu and Liu, Zhe and Liu, Xiang",
    title = "{Revealing di-pion correlations for the observed substructure near the $\pi^+\pi^-$ mass threshold in $\psi(3686)\to J/\psi \pi^+\pi^-$}",
    eprint = "2511.10345",
    archivePrefix = "arXiv",
    primaryClass = "hep-ph",
    month = "11",
    year = "2025"
}

@article{Kaiser:1998fi,
    author = "Kaiser, Norbert",
    title = "{$\pi \pi$ S-wave phase shifts and nonperturbative chiral approach}",
    doi = "10.1007/s100500050183",
    journal = "Eur. Phys. J. A",
    volume = "3",
    pages = "307--309",
    year = "1998"
}

@article{Hoferichter:2012wf,
    author = "Hoferichter, M. and Ditsche, C. and Kubis, B. and Mei{\ss}ner, U.-G.",
    title = "{Dispersive analysis of the scalar form factor of the nucleon}",
    eprint = "1204.6251",
    archivePrefix = "arXiv",
    primaryClass = "hep-ph",
    doi = "10.1007/JHEP06(2012)063",
    journal = "JHEP",
    volume = "06",
    pages = "063",
    year = "2012"
}

@article{Briceno:2017qmb,
    author = "Brice\~no, Raul A. and Dudek, Jozef J. and Edwards, Robert G. and Wilson, David J.",
    title = "{Isoscalar $\pi\pi, K\bar{K}, \eta\eta$ scattering and the $\sigma, f_0, f_2$ mesons from QCD}",
    eprint = "1708.06667",
    archivePrefix = "arXiv",
    primaryClass = "hep-lat",
    reportNumber = "JLAB-THY-17-2534",
    doi = "10.1103/PhysRevD.97.054513",
    journal = "Phys. Rev. D",
    volume = "97",
    number = "5",
    pages = "054513",
    year = "2018"
}

@article{Cleven:2011gp,
    author = "Cleven, Martin and Guo, Feng-Kun and Hanhart, Christoph and Mei{\ss}ner, Ulf-G.",
    title = "{Bound state nature of the exotic $Z_b$ states}",
    eprint = "1107.0254",
    archivePrefix = "arXiv",
    primaryClass = "hep-ph",
    doi = "10.1140/epja/i2011-11120-6",
    journal = "Eur. Phys. J. A",
    volume = "47",
    pages = "120",
    year = "2011"
}

@article{Gasser:1984gg,
    author = "Gasser, J. and Leutwyler, H.",
    title = "{Chiral Perturbation Theory: Expansions in the Mass of the Strange Quark}",
    reportNumber = "CERN-TH-3798",
    doi = "10.1016/0550-3213(85)90492-4",
    journal = "Nucl. Phys. B",
    volume = "250",
    pages = "465--516",
    year = "1985"
}

@article{Guo:2004dt,
    author = "Guo, F.-K. and Shen, P.-N. and Chiang, H.-C. and Ping, R.-G.",
    title = "{Heavy quarkonium $\pi^+ \pi^-$ transitions and a possible $b \bar b q \bar q$ state}",
    eprint = "hep-ph/0410204",
    archivePrefix = "arXiv",
    doi = "10.1016/j.nuclphysa.2005.07.019",
    journal = "Nucl. Phys. A",
    volume = "761",
    pages = "269--282",
    year = "2005"
}

@article{Guo:2006ya,
    author = "Guo, Feng-Kun and Shen, Peng-Nian and Chiang, Huan-Ching",
    title = "{Chromo-polarizability and $\pi \pi$ final state interaction}",
    eprint = "hep-ph/0604252",
    archivePrefix = "arXiv",
    doi = "10.1103/PhysRevD.74.014011",
    journal = "Phys. Rev. D",
    volume = "74",
    pages = "014011",
    year = "2006"
}

@article{Kubis:2015sga,
    author = "Kubis, Bastian and Plenter, Judith",
    title = "{Anomalous decay and scattering processes of the $\eta $ meson}",
    eprint = "1504.02588",
    archivePrefix = "arXiv",
    primaryClass = "hep-ph",
    doi = "10.1140/epjc/s10052-015-3495-5",
    journal = "Eur. Phys. J. C",
    volume = "75",
    number = "6",
    pages = "283",
    year = "2015"
}

@article{Kang:2013jaa,
    author = "Kang, Xian-Wei and Kubis, Bastian and Hanhart, Christoph and Mei{\ss}ner, Ulf-G.",
    title = "{$B_{l4}$ decays and the extraction of $|V_{ub}|$}",
    eprint = "1312.1193",
    archivePrefix = "arXiv",
    primaryClass = "hep-ph",
    doi = "10.1103/PhysRevD.89.053015",
    journal = "Phys. Rev. D",
    volume = "89",
    pages = "053015",
    year = "2014"
}

@article{Belle:2013yex,
    author = "Liu, Z. Q. and others",
    collaboration = "Belle",
    title = "{Study of $e^+e^- \to \pi^+ \pi^- J/\psi$ and Observation of a Charged Charmoniumlike State at Belle}",
    eprint = "1304.0121",
    archivePrefix = "arXiv",
    primaryClass = "hep-ex",
    reportNumber = "BELLE-PREPRINT-2013-6, KEK-PREPRINT-2013-2",
    doi = "10.1103/PhysRevLett.110.252002",
    journal = "Phys. Rev. Lett.",
    volume = "110",
    pages = "252002",
    year = "2013",
    note = "[Erratum: Phys. Rev. Lett. \textbf{111}, 019901 (2013)]"
}

@article{Chen:2023def,
    author = "Chen, Yun-Hua and Du, Meng-Lin and Guo, Feng-Kun",
    title = "{Precise determination of the pole position of the exotic $Z_{c}(3900)$}",
    eprint = "2310.15965",
    archivePrefix = "arXiv",
    primaryClass = "hep-ph",
    doi = "10.1007/s11433-023-2408-1",
    journal = "Sci. China Phys. Mech. Astron.",
    volume = "67",
    number = "9",
    pages = "291011",
    year = "2024"
}

@article{Chen:2016mjn,
    author = "Chen, Yun-Hua and Cleven, Martin and Daub, Johanna T. and Guo, Feng-Kun and Hanhart, Christoph and Kubis, Bastian and Mei\ss{}ner, Ulf-G. and Zou, Bing-Song",
    title = "{Effects of $Z_b$ states and bottom meson loops on $\Upsilon(4S) \to \Upsilon(1S,2S) \pi^+\pi^-$ transitions}",
    eprint = "1611.00913",
    archivePrefix = "arXiv",
    primaryClass = "hep-ph",
    doi = "10.1103/PhysRevD.95.034022",
    journal = "Phys. Rev. D",
    volume = "95",
    number = "3",
    pages = "034022",
    year = "2017"
}

@article{Garcia-Martin:2011iqs,
    author = "Garc\'ia-Mart\'in, R. and Kami\'nski, R. and Pel\'aez, J. R. and Ruiz de Elvira, J. and Yndur\'ain, F. J.",
    title = "{The Pion--pion scattering amplitude. IV: Improved analysis with once subtracted Roy-like equations up to 1100 MeV}",
    eprint = "1102.2183",
    archivePrefix = "arXiv",
    primaryClass = "hep-ph",
    doi = "10.1103/PhysRevD.83.074004",
    journal = "Phys. Rev. D",
    volume = "83",
    pages = "074004",
    year = "2011"
}

@article{Baru:2020ywb,
    author = "Baru, V. and Epelbaum, E. and Filin, A. A. and Hanhart, C. and Mizuk, R. V. and Nefediev, A. V. and Ropertz, S.",
    title = "{Insights into $Z_b(10610)$ and $Z_b(10650)$ from dipion transitions from $\Upsilon(10860)$}",
    eprint = "2012.05034",
    archivePrefix = "arXiv",
    primaryClass = "hep-ph",
    doi = "10.1103/PhysRevD.103.034016",
    journal = "Phys. Rev. D",
    volume = "103",
    number = "3",
    pages = "034016",
    year = "2021"
}

@article{Anisovich:1996tx,
    author = "Anisovich, A. V. and Leutwyler, H.",
    title = "{Dispersive analysis of the decay $\eta \to 3 \pi$}",
    eprint = "hep-ph/9601237",
    archivePrefix = "arXiv",
    reportNumber = "BUTP-95-2",
    doi = "10.1016/0370-2693(96)00192-X",
    journal = "Phys. Lett. B",
    volume = "375",
    pages = "335--342",
    year = "1996"
}

@article{Chen:2015jgl,
    author = "Chen, Yun-Hua and Daub, Johanna T. and Guo, Feng-Kun and Kubis, Bastian and Mei\ss{}ner, Ulf-G. and Zou, Bing-Song",
    title = "{Effect of $Z_b$ states on $\Upsilon(3S)\to\Upsilon(1S)\pi\pi$ decays}",
    eprint = "1512.03583",
    archivePrefix = "arXiv",
    primaryClass = "hep-ph",
    doi = "10.1103/PhysRevD.93.034030",
    journal = "Phys. Rev. D",
    volume = "93",
    number = "3",
    pages = "034030",
    year = "2016"
}

@article{Omnes:1958hv,
    author = "Omn{\`e}s, R.",
    title = "{On the Solution of certain singular integral equations of quantum field theory}",
    doi = "10.1007/BF02747746",
    journal = "Nuovo Cim.",
    volume = "8",
    pages = "316--326",
    year = "1958"
}

@article{Chen:2019gty,
    author = "Chen, Yun-Hua and Guo, Feng-Kun",
    title = "{Chromopolarizabilities of bottomonia from the $\Upsilon(2S,3S,4S) \to \Upsilon(1S,2S)\pi\pi$ transitions}",
    eprint = "1906.05766",
    archivePrefix = "arXiv",
    primaryClass = "hep-ph",
    doi = "10.1103/PhysRevD.100.054035",
    journal = "Phys. Rev. D",
    volume = "100",
    number = "5",
    pages = "054035",
    year = "2019"
}

@article{Dai:2014lza,
    author = "Dai, Ling-Yun and Pennington, M. R.",
    title = "{Two photon couplings of the lightest isoscalars from BELLE data}",
    eprint = "1403.7514",
    archivePrefix = "arXiv",
    primaryClass = "hep-ph",
    reportNumber = "JLAB-THY-14-1869",
    doi = "10.1016/j.physletb.2014.07.005",
    journal = "Phys. Lett. B",
    volume = "736",
    pages = "11--15",
    year = "2014"
}

@article{Brown:1975dz,
  title = {Chiral {{Symmetry}} and {$\psi$}{\textsuperscript{{$\prime$}}}{$\rightarrow\psi\pi\pi$} {{Decay}}},
  author = {Brown, Lowell S. and Cahn, Robert N.},
  year = 1975,
  journal = {Phys. Rev. Lett.},
  volume = {35},
  pages = {1},
  doi = {10.1103/PhysRevLett.35.1}
}

@article{Pelaez:2015qba,
  title = {From controversy to precision on the sigma meson: {{A}} review on the status of the non-ordinary {{$f_0(500)$}} resonance},
  author = {Pel{\'a}ez, J. R.},
  year = 2016,
  journal = {Phys. Rept.},
  volume = {658},
  eprint = {1510.00653},
  primaryclass = {hep-ph},
  pages = {1},
  doi = {10.1016/j.physrep.2016.09.001},
  archiveprefix = {arXiv}
}

@article{Anisovich:1995zu,
  title = "{$\Upsilon(3S)\to \Upsilon(1S)\pi\pi$ decay: Is the $\pi\pi$ spectrum puzzle an indication of a $b\bar b q\bar q$ resonance?}",
  author = {Anisovich, V. V. and Bugg, D. V. and Sarantsev, A. V. and Zou, B.-S.},
    doi = "10.1103/PhysRevD.51.R4619",
    journal = "Phys. Rev. D",
    volume = "51",
    pages = "R4619--R4622",
    year = "1995"
}

@article{Guo:2006ai,
  title = "{On the structure of the $\pi\pi$ invariant mass spectra of the $\Upsilon(4S) \to \Upsilon(1S,2S) \pi\pi$ decays}",
  author = {Guo, Feng-Kun and Shen, Peng-Nian and Chiang, Huan-Ching and Ping, Rong-Gang},
  year = 2007,
  journal = {Phys. Lett. B},
  volume = {658},
  eprint = {hep-ph/0601120},
  pages = {27--32},
  doi = {10.1016/j.physletb.2007.10.021},
  archiveprefix = {arXiv}
}

@article{Sommerfeld:1931qaf,
  title = {{\"U}ber die {{Beugung}} und {{Bremsung}} der {{Elektronen}}},
  author = {Sommerfeld, A.},
  year = 1931,
  journal = {Annalen Phys.},
  volume = {403},
  pages = {257--330},
  doi = {10.1002/andp.19314030302}
}

@article{Yan:1998wv,
    author = "Yan, Mu-Lin and Wei, Yi and Zhuang, Ting-Liang",
    title = "{Comment on the hadronic decay of excited heavy quarkonia}",
    eprint = "hep-ph/9805354",
    archivePrefix = "arXiv",
    doi = "10.1007/s100520050384",
    journal = "Eur. Phys. J. C",
    volume = "7",
    pages = "61--63",
    year = "1999"
}

@article{Yan:2026oil,
    author = "Yan, Jiang and Cao, Xiong-Hui and Du, Meng-Lin and Guo, Feng-Kun",
    title = "{Scattering lengths of the $J/\psi\pi$ and $J/\psi K$ systems}",
    eprint = "2601.18103",
    archivePrefix = "arXiv",
    primaryClass = "hep-ph",
    month = "1",
    year = "2026",
    journal = ""
}

\end{document}